\documentclass[12pt]{iopart}
\begin{document}
\jl{4}
\title[Radiative electroweak effects in polarized DIS ]
{Radiative electroweak effects in deep inelastic scattering
of polarized leptons by polarized nucleons}
\author {I. V. Akushevich\dag, A.N. Ilyichev\dag\ and N. M.
Shumeiko \dag}
\address{\dag\  National Scientific and Education Center of Particle and
 High Energy Physics attached to Byelorussian State University}
\begin{abstract}
The one-loop electroweak radiative correction of the
 lowest order to lepton current for deep inelastic scattering (DIS) of
longitudinally polarized leptons by polarized 
nucleons is obtained  in model independent way.
The detailed numerical analysis within  kinematical requirements of 
future polarization collider experiments is performed. The possibility to
reduce radiative effects by detection of hard photons is studied.
\end{abstract}
\submitted
\maketitle

\input{epsf}

\section{Introduction}
Deep inelastic lepton-hadron scattering, starting with the
discovery of Bjorken scaling in the end of nineteen-sixties has
played a crucial role in the development of our present
understanding of the nature of particle interaction. The
appearance of the first data on the polarization DIS
opened a new field of experimental and theoretical investigations
and, as a result it led to come the disturbing data of EMC in 1988
\cite{EMC}.

It is natural, that data processing of the modern experiments on DIS of
the polarized lepton on the polarized nuclear target requires correct
account of the radiative corrections (RC).  Up to now there is a series
of works where RC are taken into account in the frame of QED-theory for
polarization experiments (see f.e.  \cite{KSh,ASh,SSh,Tim}). Those results
are used for experiments on a fixed target.  However the polarization DIS
experiments at collider will be possible in future \cite{HERA}. 

\begin{figure} 
\vspace{1cm}
\begin{tabular}{ccccc} 
\begin{picture}(60,100)  
\put(-80,-110){
\epsfxsize=8cm 
\epsfysize=9cm 
\epsfbox{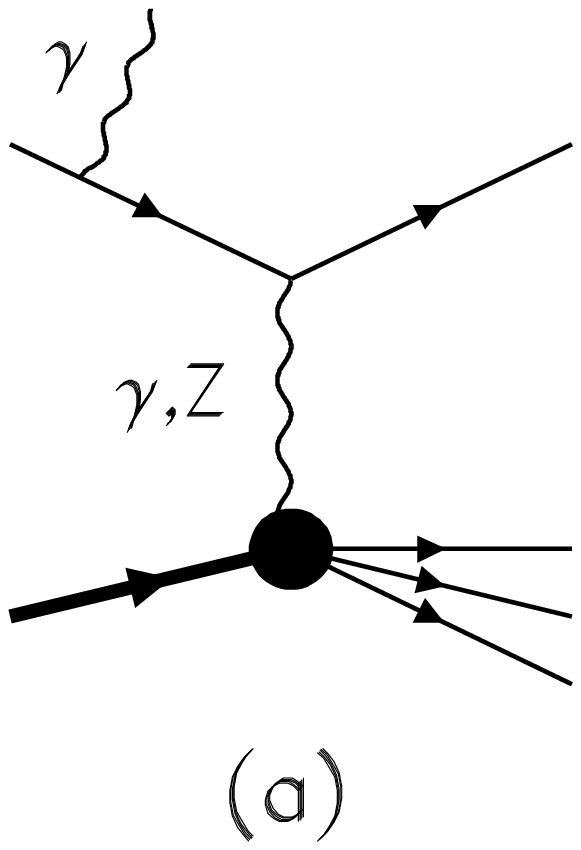} } 
\end{picture} 
&
\begin{picture}(60,100)  
\put(-60,-110){ 
\epsfxsize=8cm 
\epsfysize=9cm
\epsfbox{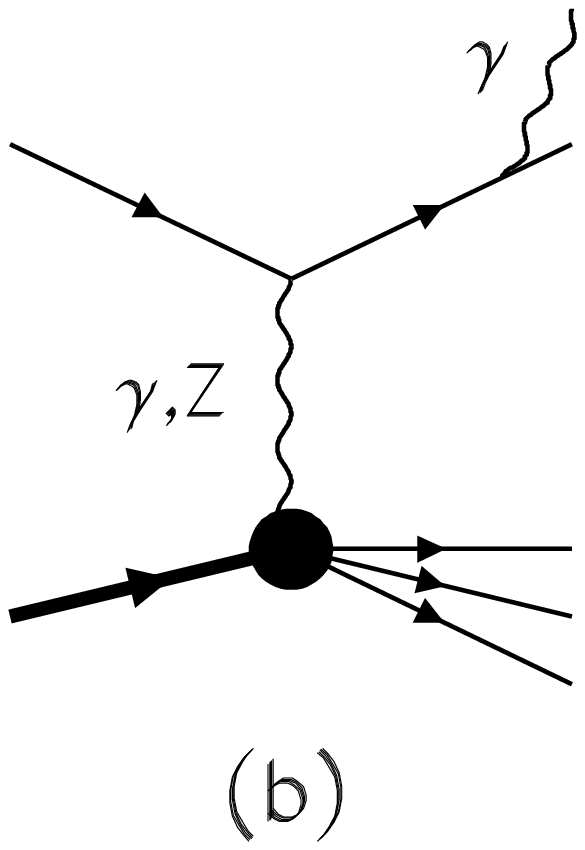} } 
\end{picture} 
& 
\begin{picture}(60,100) 
\put(-40,-110){ 
\epsfxsize=8cm 
\epsfysize=9cm 
\epsfbox{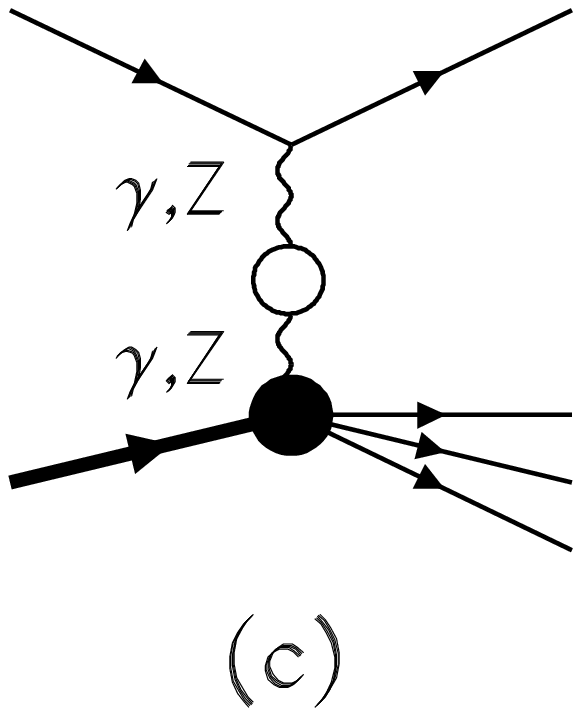} }
\end{picture} 
& 
\begin{picture}(60,100)  
\put(-20,-110){ 
\epsfxsize=8cm
\epsfysize=9cm 
\epsfbox{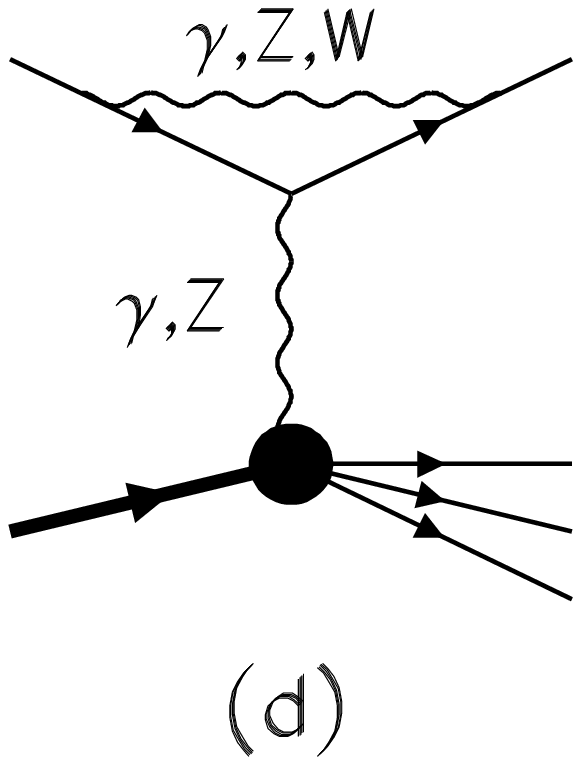} } 
\end{picture} 
&
\begin{picture}(60,100)  
\put(0,-110){ 
\epsfxsize=8cm 
\epsfysize=9cm
\epsfbox{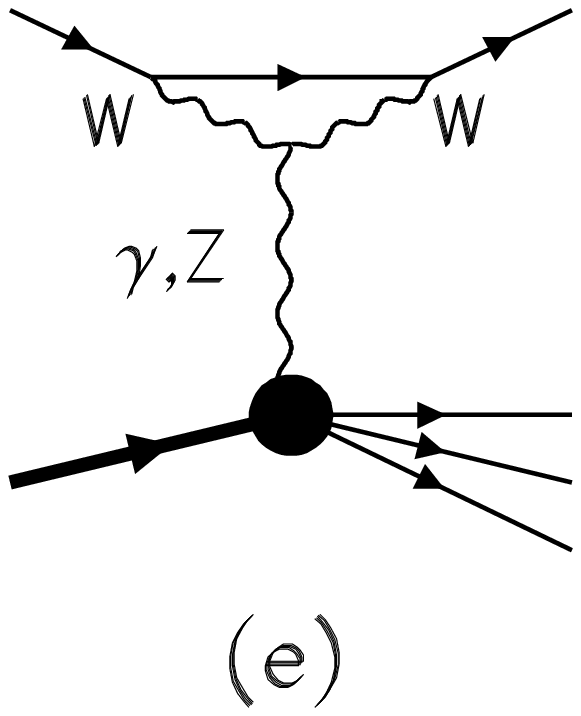} } 
\end{picture} 
\end{tabular} 
\caption{\it Full set of Feynman graphs  contributing to the model independent 
electroweak correction of the  lowest order. All possible
graphs, which give the contribution to vacuum polarization,
are designated by a symbol of an open circle \cite{Holl}.
} 
\label{feyn} 
\end{figure}

For the calculation of RC for such kinds of experiments we cannot restrict
our consideration to $\gamma$-exchange graphs only, because in this case
the squared transfer momentum $Q^2$ is so high that weak effects begin to
play an essential role in the total cross section and spin asymmetries.
At the same time the ratio $m^2/Q^2$ (where $m$ is the mass of a
scattering
lepton) becomes so small that it could be restricted to non-vanishing
terms for $m\rightarrow 0$.  The similar work was already done by us
\cite{AISh}, but there we used only the naive parton model. In this
article we present the explicit expressions for model independent part of
the one-loop lowest-order RC (figure \ref{feyn})
to polarized lepton-nucleon scattering which is described in the terms of
the electroweak structure functions (SF).  Moreover, the target has
an arbitrary polarization and the lepton has a longitudinal one. The
contributions appearing from additional virtual particles (V-contribution)
in the on-mass renormalization scheme and t'Hooft-Feynman gauge are
presented.  The results of calculation for the unitary gauge can be found
in
ref.  \cite{Bardin}. We note also that the separation of variables in
accordance with \cite{ASh} allows to write all formulae in a compact form
that provided more clearness of the results than it was done in
\cite{Bardin}.

The present article is organized as follows. In the section \ref{born} the
Born contribution and all necessary kinematic invariants are presented.
The section 3 is devoted to the electroweak one-loop correction.  Numerical
analysis for kinematical conditions of the collider experiments is
presented in the section 4. Conclusion is given in the last section. 

\section{Born contribution }
\label{born}
Here we consider the deep inelastic polarized lepton-hadron
scattering
\begin{equation}
\ell (k_{1},\xi ) + N(p,\eta ) \rightarrow  \ell (k_{2}) + X
\label{proc}
\end{equation}
in the frame of electroweak standard theory.
The quantities in brackets  $k_1 (k_2)$, $p$ define the momenta of
an initial (final) lepton and proton
respectively ($k_1^2=k_2^2=m^2$, $p^2=M^2$), $\xi $ and $\eta $ are the
polarization vectors of the scattering particles.

Since the lepton is considered to be longitudinally polarized, its
polarization vector has the form \cite{ASh}:
\begin{equation}
\xi  = {S\over m \sqrt{\lambda _{s}}}k_{1}-{2m\over \sqrt{\lambda_{s}}}p
=\xi _0+\xi '.
\label{xi}
\end{equation}
The kinematical invariants  are defined in a standard way:

\begin{equation}
\eqalign {
 S = 2k_{1}p,\; X = 2k_{2}p = (1-y)S,\;
Q^2=-q^2=-(k_{1}-k_{2})^{2}={ xyS},
\\
 S_{x}= S-X,\quad S_p=S+X,\quad 
\lambda _{s}=S^{2}-4m^{2}M^{2},}
\label{invar}
\end{equation}
where $x$ and $y$ are usual scaling variables.

The double-differential cross section of lepton-nucleon
scattering on the Born level 
($d\sigma ^B /dxdy \equiv \sigma^B$) 
in the frame of electroweak theory reads
\begin{equation}
\fl \sigma ^B= {4\pi \alpha ^{2} S_{x}S\over \lambda
 _{s}Q^4
}\left[
L^{\gamma\gamma}_{\mu \nu}W_{\mu \nu}^{\gamma}(p,q)
+\frac 12(L^{\gamma Z}_{\mu \nu}
+L^{Z\gamma }_{\mu \nu})W_{\mu \nu}^{\gamma Z}(p,q)\chi
+L^{ZZ}_{\mu \nu}W_{\mu \nu}^{Z}(p,q)\chi ^2
\right ],
\label {b1}
\end {equation}
where
$\chi=Q^2/(Q^2+M_Z^2)$ ($M_Z$ is the Z-boson mass).
The lepton tensors
$ L^{mn}_{\mu \nu}$ can be given as

\begin{equation}
\fl
L^{mn }_{\mu \nu}=\frac 14 Sp \; \gamma _{\mu }(v^m-a^m\gamma _5)(\hat
{k}_1+m)(1-P_L\gamma_5 \hat {\xi})\gamma_{\nu}(v^n-a^n\gamma _5)
(\hat {k}_2+m),
\label {l1}
\end {equation}
where
\begin{equation}
\eqalign {
v^{\gamma }=1,\qquad&
v^Z=(-1+4s_w ^2)/4s_w c_w,
\\
a^{\gamma }=0,&
a^Z=-1/4s_wc_w
}
\end{equation}
are the standard electroweak coupling constants, $c_w$ and $s_w$ are 
cosin and sine of Weinberg's angle respectively
 and $P_L$ is degree of
lepton
polarization.

All of the hadronic tensors $W^{\gamma ,\gamma Z,Z}_{\mu \nu}(p,q)$ can be
expressed in terms of eight electroweak SF  \cite{DIS, Bl}:
\begin{eqnarray}
\fl W^I_{\mu \nu}(p,q)=\sum_{i=1}^8
w^i_{\mu \nu}{\bar F}^I_i=
-\tilde{g}_{\mu \nu}\bar{F}_1^I
+\frac 1{M^2}\tilde{p}_{\mu }\tilde{p}_{\nu}\bar{F}_2^I \nonumber
\\
\lo
+i\epsilon_{\mu \nu \lambda \sigma } \Biggl[\frac {p^{\lambda
}q^{\sigma }}{2M^2}
\bar{F}_3^I
+q^{\lambda }\eta^{\sigma
} \frac 1M
\bar{F}^I_4
-q^{\lambda }p^{\sigma }
\frac{\eta q}{M^3}\bar{F}^I_5\Biggl] \nonumber
\\
\lo
-\frac 1{2M}\left[
\tilde{p}_{\mu }\tilde{\eta}_{\nu }+
\tilde{p}_{\nu }
\tilde{\eta} _{\mu }
\right]\bar{F}^I_6
+\frac {\eta q}{M^3}\tilde{p}_{\mu }
\tilde{p}_{\nu}\bar{F}^I_7
+\frac{\eta q}{M}\tilde{g}_{\mu \nu}\bar{F}^I_8
\label{w}
\end{eqnarray}
where $I=\gamma,\gamma Z,Z$,
${\bar F}$ are defined as ($\epsilon=M^2/pq$)
\begin{equation}
\fl 
{\bar F}^I_1=F^I_1,\;
{\bar F}^I_{2,3}=\epsilon F^I_{2,3},\;
{\bar F}^I_{4}=\epsilon P_N(g^I_1+g^I_2),\;
{\bar F}^I_{5,7}=\epsilon^2 P_Ng^I_{2,4},\;
{\bar F}^I_{6,8}=\epsilon P_Ng^I_{3,5}
\label{sf}
\end{equation}
and 
\begin{equation}
\tilde{g}_{\mu\nu}= g_{\mu \nu}+\frac {q_{\mu }q_{\mu }}{Q^2},\;\;\;
\tilde{p}_{\nu}=p_{\nu}+\frac{pq}{Q^2}q_{\nu },\;\;\;
\tilde{\eta}_{\nu}=\eta_{\nu}+\frac{\eta q}{Q^2}q_{\nu }.
\label{td}
\end{equation}
Hadronic tensors in the form (\ref{w} - \ref{td}) correspond to the
definition of electroweak SF given in (2.2.7) of ref. \cite{DIS}.
For contraction of leptonic tensor (\ref{l1}) with $w^i_{\mu \nu}$
we have
\begin{equation}
\eqalign {
L^{mn}_{\mu \nu}w^i_{\mu \nu}=\theta^B_i R^{mn}_V\;\;&
(i=1,2,6-8),
\\
L^{mn}_{\mu \nu}w^i_{\mu \nu}=\theta^B_i R^{mn}_A&
(i=3-5).
}
\label{contb}
\end{equation}
The quadratic combinations of the electroweak coupling constants are
defined as:
\begin{eqnarray}
\eqalign {
 R^{mn}_V=(v^mv^n+a^ma^n)-P_L(v^ma^n+v^na^m),\;\;
\\
 R^{mn}_A=(v^ma^n+a^mv^n)-P_L(v^mv^n+a^na^m).
}
\end{eqnarray}
The quantities $\theta^B_i$
depend only on the target polarization vector
and kinematical invariants:
\begin{equation}
\eqalign{
\theta^B_1=Q^2,&
\theta^B_5=\eta q Q^2S_p/2M^3,\\
\theta^B_2=(SX-M^2Q^2)/2M^2,\;\;&
\theta^B_6=-(X\eta k_1+S\eta k_2)/2M,\\
\theta^B_3=Q^2S_p/4M^2,&
\theta^B_7=\eta q(SX-M^2Q^2)/2M^3,\\
\theta^B_4=-Q^2\eta ( k_1+k_2)/M,&
\theta^B_8=-\eta qQ^2/M.}
\label{tb}
\end{equation}

Covariant representation for the proton polarization vector both for
longitudinally and transversely polarized nucleons $\eta $ can be found in
Appendix A of ref.\cite{ASh}. It does not contain  a nucleon polarization
degree $P_N$ which is included in SF definition (\ref{sf}). 

 We note, that only leading part of $\xi $ ($\xi _0$) contributes to
$\theta^B_i$ in the ultrarelativistic approximation. The second term ($\xi
'$) 
gives non-vanishing correction to the cross section of radiated process
which is considered below. 

 Using the hadronic tensor (\ref{w}) and the results for contractions
(\ref{contb}) the cross section (\ref{b1}) can be rewritten in a simple
form

\begin{equation}
\sigma ^B={4\pi\alpha ^{2}y\over Q^4}
\sum^{8}_{i=1}
\theta^B_i{\cal F}_{i}.
\label {b2}
\end {equation}

Both for  the Born cross section and for RC considered in the next section
the electroweak SF are gathered in eight combinations.
So it is convenient to define the generalized SF:
 
\begin{equation}
\eqalign{
{\cal F}_i=R^{\gamma}_V{\bar F}^{\gamma }_i+
\chi R^{\gamma Z}_V{\bar F}^{\gamma Z}_i+
\chi^2 R^Z_V{\bar F}^Z_i\;\;&
(i=1,2,6-8),\\
{\cal F}_i=R^{\gamma}_A{\bar F}^{\gamma }_i+
\chi R^{\gamma Z}_A{\bar F}^{\gamma Z}_i+
\chi^2 R^Z_A{\bar F}^Z_i&
(i=3-5).
}
\label {calf}
\end{equation}

\section {The Lowest Order Radiative Correction }

The RC of the lowest order appears as the result of one-loop effects
and the process with a real photon radiation:
\begin{equation}
\ell (k_{1},\xi ) + N(p,\eta ) \rightarrow  \ell (k_{2}) +
\gamma(k)+ X.
\label{rproc}
\end{equation}
The cross section of the radiated process 
($d\sigma^R /dxdy \equiv \sigma^R$) 
can be presented as the sum of
two parts:
\begin{equation}
 \sigma ^R=
\bar{\sigma} ^R+
\hat{\sigma} ^R.
\label {r}
\end{equation}
The first one ($\bar{\sigma} ^R$) includes
a part of the cross section independent of the leptonic polarization
vector (\ref{xi}) and contribution of its leading term $\xi_0$.
 The second term ($\hat{\sigma} ^R$) comes from $\xi '$ only.
We can write them again in the terms of leptonic and hadronic
tensors:
\begin{eqnarray}
\fl \bar{\sigma} ^R=- {\alpha ^{3}S_{x}S\over \pi \lambda
_{s}}\int {d^3k\over k_0}
{1\over Q_h^4}\Biggl [
\bar{\cal L}^{\gamma \gamma}_{\mu \nu}W_{\mu \nu}^{\gamma}(p,q_h)
+\frac12(
\bar{\cal L}^{\gamma Z}_{\mu \nu}+
\bar{\cal L}^{Z \gamma }_{\mu \nu})
W_{\mu \nu}^{\gamma Z}(p,q_h)\chi _h \nonumber
\\ 
\lo+\bar{\cal L}^{Z}_{\mu \nu}W_{\mu \nu}^{Z}(p,q_h)\chi _h^2
\Biggl ],
\label {r1}
\end {eqnarray}
\begin{eqnarray}
\fl \hat{\sigma} ^R=- {\alpha ^{3}S_{x}S\over \pi \lambda
 _{s}}\int\ {d^3k\over k_0}
{1\over Q_h^4}\Biggl [
\hat{\cal L}^{\gamma \gamma}_{\mu \nu}W_{\mu \nu}^{\gamma}(p,q_h)
+\frac12(
\hat{\cal L}^{\gamma Z}_{\mu \nu}+
\hat{\cal L}^{Z \gamma }_{\mu \nu})
W_{\mu \nu}^{\gamma Z}(p,q_h)\chi _h \nonumber
\\ 
\lo+\hat{\cal L}^{Z}_{\mu \nu}W_{\mu \nu}^{Z}(p,q_h)\chi _h^2
\Biggl].
\label {r2}
\end {eqnarray}
Here $ \chi _h=Q_h^2/(Q_h^2+M_Z^2) $ and $
Q_h^2=-q^2_h=-(k_1-k-k_2)^2$
are the variables dependent of the momentum of a real photon $k$.

The hadronic tensors $W^{\gamma ,\gamma Z,Z}_{\mu
\nu}(p,q_h)$ are defined by (\ref{w}). The leptonic tensors
in (\ref{r1})  include spin averaged and leading spin dependent 
parts
\begin{equation}
 \bar {\cal L}^{mn}_{\mu \nu}=\frac 14 Sp \; \Gamma ^m_{\mu \alpha}
(\hat {k}_1+m)(1-P_L\gamma_5 \hat{\xi}_0)
\bar{\Gamma }^n_{\alpha \nu}(\hat {k}_2+m),
\label {cl1}
\end {equation}
and $\hat{\sigma} ^R$ (\ref{r2})  
comes from the contribution of $\xi ' $
only:
\begin{equation}
 {\hat {\cal L}}^{mn}_{\mu \nu}=P_L\frac 14 Sp \;
\Gamma ^m_{\mu \alpha}(\hat {k}_1+m)\gamma_5
\hat {\xi}'\bar{\Gamma} ^n_{\alpha \nu}
(\hat {k}_2+m),
\label {cl2}
\end {equation}
where
\begin {equation}
\eqalign{
 \Gamma^m _{\mu \alpha}=
\left [\left(\frac {k_{1 \alpha}}{kk_1}- \frac {k_{2 \alpha}}{kk_2}
\right)\gamma_{\mu}-
\frac {\gamma_{\mu}\hat{k}\gamma_{\alpha}}{2kk_1}-
\frac{\gamma_{\alpha}\hat{k}\gamma_{\mu}}{2kk_2} \right](v^m-a^m\gamma
_5),
\\
 \bar{\Gamma}^n _{\alpha \nu}=
\left [\left(\frac {k_{1 \alpha}}{kk_1}- \frac {k_{2 \alpha}}{kk_2}
\right)\gamma_{\nu}
-\frac{\gamma_{\alpha}\hat{k}\gamma_{\nu}}{2kk_1}
-\frac {\gamma_{\nu}\hat{k}\gamma_{\alpha}}{2kk_2}
 \right](v^n-a^n\gamma
_5),
}
\label{g12}
\end {equation}

The results for the  contraction of these tensors with $w^i_{\mu \nu }$
(see (\ref{w})) can be presented using notations of 
ref.\cite{ASh,P20}:
\begin{equation}
\eqalign{
\frac 1{\pi} \int \frac {d^3k}{k_0}
\bar{\cal L}^{mn}_{\mu \nu}w^i_{\mu \nu}=
R^{mn}_V
\int dR d\tau
\sum_{j=1}^{k_i}R^{j-2}{\theta}_{ij}(\tau)
\; \; &
(i=1,2,6-8),
\\
\frac 1{\pi} \int \frac {d^3k}{k_0}
\bar{\cal L}^{mn}_{\mu \nu}w^i_{\mu \nu}=
R^{mn}_A
\int dR d\tau
\sum_{j=1}^{k_i}R^{j-2}{\theta}_{ij}(\tau)
&
(i=3-5),
}
\end{equation}
where $j$ runs from $1$ to $k_i=(3,3,4,4,5,3,4,4)$
and the quantities ${\theta}_{ij}(\tau )$ are independent of $R$.
Arguments of SF $x$ and $Q^2$ in the case of radiated process acquire
dependence on two photonic variables $R=2kp$ and $\tau=2kq_h/R $: 
\begin{equation}
Q^2\rightarrow Q^2+R\tau, \qquad
x\rightarrow {Q^2+R\tau \over S_x-R}.
\label{arg}
\end{equation}
Integration over third photonic variable is performed analytically.
Then the cross section $\bar{\sigma} ^R$ can be obtained in the form
\begin {equation}
\bar{\sigma} ^R= - \alpha ^{3}y
\int\limits^{\tau_{max}}_{\tau_{min}} d\tau\sum^{8}_{i=1}
 \sum^{k_{i}}_{j=1} \theta _{ij}(\tau)\int \limits^{R_{max}}_{0} dR
 {R^{j-2}\over (Q^2+R\tau)^{2}}{\cal F}_{i}(R,\tau),
\label {in}
\end {equation}
where integration limits are:
\begin{equation}
\tau _{max,min} = {S_{x} \pm \sqrt {S_x^2+4M^2Q^2}\over 2M^{2}},\quad
R_{max} ={{W^{2}-(M+m_{\pi })^2}\over 1+\tau}.
\end {equation}
Here $m_{\pi}$ is the pion mass  and $W^{2}=S_{x}-Q^2+M^{2}$
is the squared mass of final hadrons.
Summing up over $i=1,...,8$ corresponds to the contribution of the
generalized SF ${\cal F}_{i}(R, \tau)$ defined by
(\ref{calf}) with the  replacement of arguments (\ref{arg}).
The infrared divergence occurs in the integral for $R\rightarrow 0$ in the
term where $j=1$ (and only in it).
Explicit expressions for $\theta _{ij}(\tau )$ are given in Appendix.

The cross section $\hat{\sigma} ^R$ can be found
in the  following way. For the contraction we have
\begin{equation}
\eqalign{
\fl
\frac 1{\pi} \int \frac {d^3k}{k_0}
\hat{\cal L}^{mn}_{\mu \nu}w^i_{\mu \nu}=
P_L(v^na^m+a^nv^m)
\int dR d\tau R
\hat{\theta}_{i}(R,\tau)
\frac{m^2B_1(\tau)}{C_1^{3/2}(\tau )}\;\;
(i=1,2,6-8),
\\
\fl
\frac 1{\pi} \int \frac {d^3k}{k_0}
\hat{\cal L}^{mn}_{\mu \nu}w^i_{\mu \nu}=
P_L(v^nv^m+a^na^m)
\int dR d\tau R
\hat{\theta}_{i}(R,\tau)
\frac{m^2B_1(\tau)}{C_1^{3/2}(\tau )}\;\;
(i=3-5),
}
\label{hh}
\end{equation}
where $C_1(\tau)$ and  $B_1(\tau )$ are given in Appendix (\ref{bc}).
Integrand of (\ref{hh}) has a peak coming from the region $\tau \sim \tau
_s\equiv -Q^2/S$, where $C_1(\tau) \sim m^2$. In the ultrarelativistic
approximation the integration over $\tau $ can be carried out
analytically
\begin{eqnarray}
\int\limits^{\tau_{max}}_{\tau_{min}} d\tau 
\frac {m^2B_1(\tau )}{C_1^{3/2}(\tau )}{\cal G}(\tau )
={\cal G}(\tau _s)
+\int\limits^{\tau_{max}}_{\tau_{min}} d\tau \frac{m^2B_1(\tau)}
{C_1^{3/2}(\tau )}
\left[ {\cal G}(\tau )-{\cal G}(\tau _s)\right],
\label{ts}
\end{eqnarray}
where 
\begin{equation}
\int \limits^{\tau_{max}}_{\tau_{min}}d\tau \frac {m^2B_1(\tau)}
{C_1^{3/2}(\tau )}=1
\end{equation}
was used. The quantity
\begin{equation}
{\cal G}(\tau )=\int\limits^{R_{max}}_0 \frac{RdR }{(Q^2+R\tau) ^2}  
\sum\limits^8_{i=1}\hat {\theta} _i
(R, \tau ){\cal F}_i^{pl}(R,\tau )
\end{equation}
is a function over $\tau $ regular in
the ultrarelativistic approximation.
The second term in (\ref{ts}) $\sim m^2$ and has to be dropped
in the approximation considered. The SF ${\cal F}_i^{pl}(R,\tau
)$
are parts of the generalized SF containing $P_L$. 
The quantities $\hat{\theta}_{i}(R,\tau_s)$ can be obtained from Born
ones (\ref{tb}) by the following replacements:
\begin{equation}
\hat{\theta}_{i}(R,\tau _s)=\frac 4{S(S-R)}
{\theta}_{i}^B
\left(
k_1\rightarrow \left(1-\frac R S\right) k_1 
\right)
\label{hhh}
\end{equation}
As a result the contribution $\hat \sigma_R $ can be expressed in
terms of  the Born cross section.

Below we give an explicit result for the total one-loop lowest order
correction which includes the contribution from the radiation of a 
real photon ($\sigma_R$, see figure \ref{feyn} ($a$, $b$)) and from
additional virtual particles ($\sigma_V$, see figure \ref{feyn} ($c$
- $e$))  and can be presented as the sum of four infrared free terms: 
\begin{equation}
\sigma_V+\sigma _R=
\frac{\alpha}{\pi}\delta_{VR}\sigma ^B+\sigma^r_V
+\sigma^F_R+{\hat\sigma} _R
\label {all}
\end {equation}

The factor
\begin {eqnarray}
\fl \delta _{VR}=
(\ln \frac {Q^2}{m^2}-1)\ln
\frac{(W^2-(M+m_{\pi})^2)^2}{(X+Q^2)(S-Q^2)} \nonumber
\\
\lo +\frac 32\ln \frac {Q^2}{m^2}-2-\frac 12\ln^2\frac{X+Q^2}{S-Q^2}  
+{\rm Li}_2
\frac{SX-Q^2M^2}{(X+Q^2)(S-Q^2)}
-\frac{\pi^2}6
\end {eqnarray}
appears in front of the Born cross section after
cancellation of infrared divergence by summing of
an infrared part separated from $\sigma _R$ and so-called
'QED-part' of V-contribution which arises
from the lepton vertex graphs including an additional virtual photon.

The contribution from electroweak loops with the exception of
'QED-part' can be written in terms of the Born cross section with
the following replacement:
\begin {equation}
\sigma_V^r=\sigma^ B\left(
R^{mn}_{V,A}\rightarrow \delta R^{mn}_{V,A}
\right),
\end {equation}
where
\begin{eqnarray}
\fl\delta R^{\gamma \gamma}_{V,A}=-2\Pi ^{\gamma }
R^{\gamma \gamma }_{V,A}-2\Pi ^{\gamma Z}\chi R^{Z\gamma }_{V,A}
\nonumber\\
\lo+\frac{\alpha }{4\pi}[
2R^{ZZ}_{V,A}\Lambda _2(-Q^2,M_Z)
+(1-P_L)\frac 3 {2s_w^2}\Lambda _3(-Q^2,M_W)],
\nonumber\\
\fl\delta R^{\gamma Z, Z\gamma}_{V,A}=
-2(\Pi ^{\gamma }+\Pi ^Z)
R^{\gamma Z}_{V,A}
-2\Pi ^{\gamma Z}(R^{\gamma \gamma}_{V,A}+\chi R^{ZZ}_{V,A})
\nonumber\\
\lo+\frac{\alpha }{4\pi}[2\left(v^ZR^{ZZ}_{V,A}+a^ZR^{ZZ}_{A,V}\right)
\Lambda _2(-Q^2,M_Z)
\nonumber\\
\lo+(1-P_L)\{\frac 1{8s_w^3c_w}\Lambda _2(-Q^2,M_W)
+\frac 3{4s_w^2}(v^Z+a^Z-\frac {c_w} {s_w})\Lambda
_3(-Q^2,M_W)\}],
\nonumber\\
\fl\delta R^{ZZ}_{V,A}=-2\Pi ^ZR^{ZZ}_{V,A}-2\Pi ^{\gamma
Z}R^{\gamma Z}_{V,A}
\nonumber\\
\lo+\frac{\alpha }{4\pi}[(2\left( (v^Z)^2+(a^Z)^2 \right)R^{ZZ}_{V,A}
+4v^Za^ZR^{ZZ}_{A,V})\Lambda _2(-Q^2,M_Z)
\nonumber\\
\lo+(1-P_L)(v^Z+a^Z)\{\frac 1{4s_w^3c_w}\Lambda _2(-Q^2,M_W)
-3\frac {c_w}{2s_w^3}\Lambda_3(-Q^2,M_W)\}].
\end {eqnarray}
Here $M_{Z,W}$ are the masses of $Z$ and $W$ bosons and
\begin{equation}
\begin{array}{l}
\displaystyle
\Pi ^{\gamma}=-{\hat{\Sigma }^{\gamma }(-Q^2)\over Q^2},\quad
\Pi ^Z=-{\hat{\Sigma }^Z(-Q^2)\over Q^2+M_Z^2}          ,\quad
\Pi ^{\gamma Z}=-{\hat{\Sigma }^{\gamma Z}(-Q^2)\over Q^2}.
\end{array}
\label{Pi}
\end{equation}
Quantities $\hat{\Sigma}^{\gamma, \gamma Z, Z}$ are defined by the
formulae (A.2,3.17,B.2-5) of
\cite{Holl} and $\Lambda _{2,3}$ by (B.4,B.6)  of \cite{BHS}.

The infrared free part of the cross section of the process (\ref{rproc})
has the form
\begin{eqnarray}
\fl \sigma _R^F = -\alpha ^{3}y
 \int\limits^{\tau_{max}}_{\tau_{min}}d\tau
 \sum^{8}_{i=1}
\biggl\{ \theta_{i1}(\tau)
\int\limits^{R_{max}}_{0}{dR\over R}
\left[ {{\cal F} _{i}(R,\tau )\over (Q^2+R\tau)^2}-{{\cal F}
_{i}(0,0)\over Q^4}\right]\nonumber
\\
\qquad 
\qquad 
+ \sum^{k_{i}}_{j=2}
\theta_{ij}(\tau)\int\limits^{R_{max}}_{0}
dR {R^{j-2}\over (Q^2+R\tau)^{2}}{\cal F}_{i}(R,\tau)
\biggl\}.
\label{FR}
\end{eqnarray}

As it was shown in (\ref{hh}-\ref{hhh})
the last term of (\ref{all}) 
is obtained in terms of the Born cross section 
\begin{equation}
 {\hat\sigma_ R}=\frac{\alpha y}{\pi S}\int\limits^{R^s_{max}}_0
\frac{RdR}{(S_x-R)}\tilde{\sigma}_{pl}^B,
\end{equation}
where the upper limit $R^s_{max}=S(W^2-(M+m_{\pi})^2)/(S-Q^2)$,
and $\tilde{\sigma}_{pl}^B$ is lepton polarization part of the Born 
cross section with the following replacement of kinematical variables:
$S\rightarrow
S-R$, $Q^2\rightarrow Q^2(1-R/S)$ and $k_1\eta \rightarrow k_1\eta
(1-R/S)$.

\section {Numerical Analysis }
\begin{figure}
\vspace{2cm}
\unitlength 1mm
\begin{tabular}{cc}
\begin{picture}(60,60)
\put(-5,0){
\epsfxsize=7cm
\epsfysize=8cm
\epsfbox{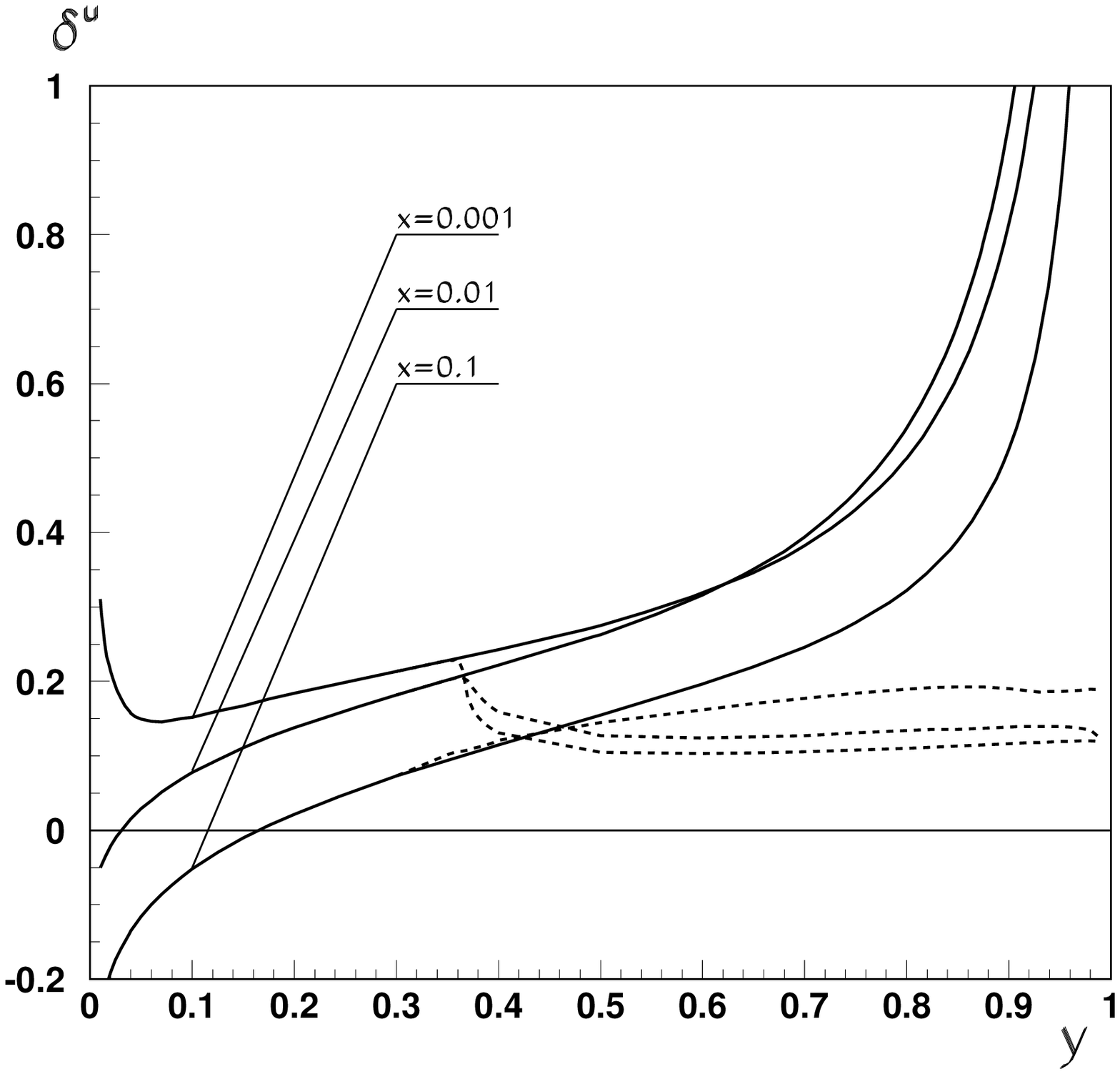}
}
\end{picture}
&
\begin{picture}(60,60)
\put(20,0){
\epsfxsize=7cm
\epsfysize=8cm
\epsfbox{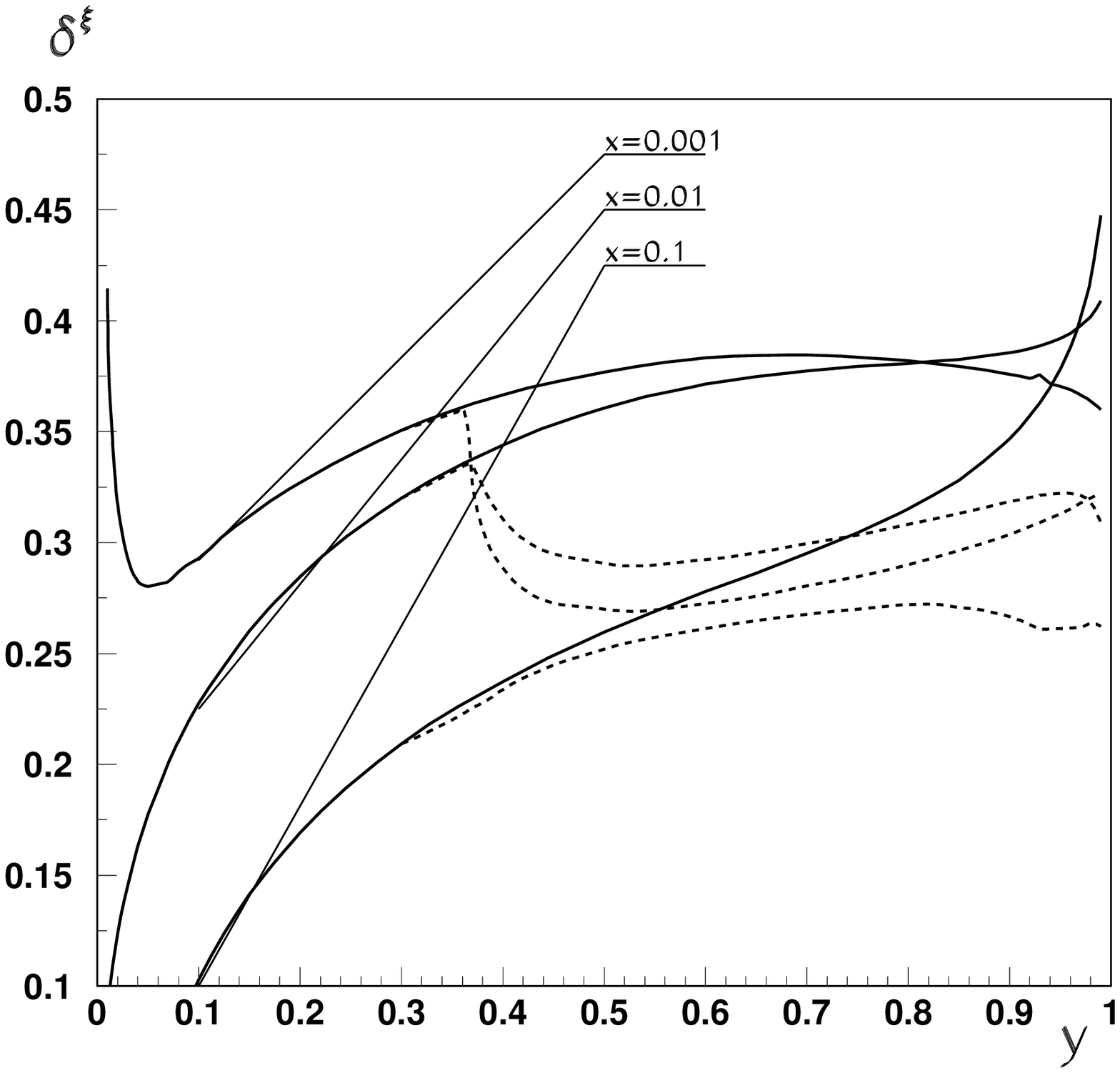}
}
\end{picture}
\\
\begin{picture}(60,60)
\put(-5,0){
\epsfxsize=7cm
\epsfysize=8cm
\epsfbox{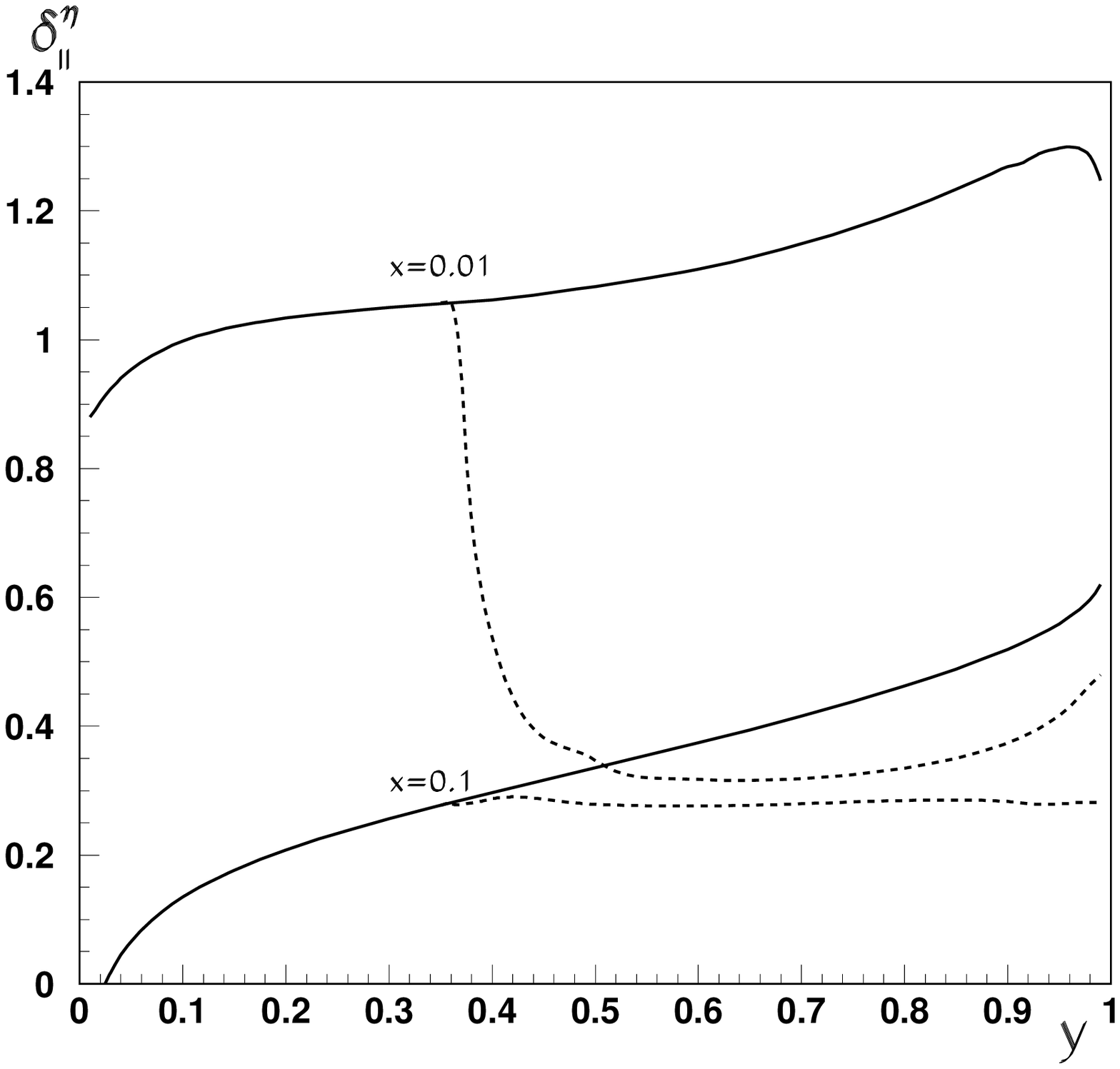}
}
\end{picture}
&
\begin{picture}(60,60)
\put(20,0){
\epsfxsize=7cm
\epsfysize=8cm
\epsfbox{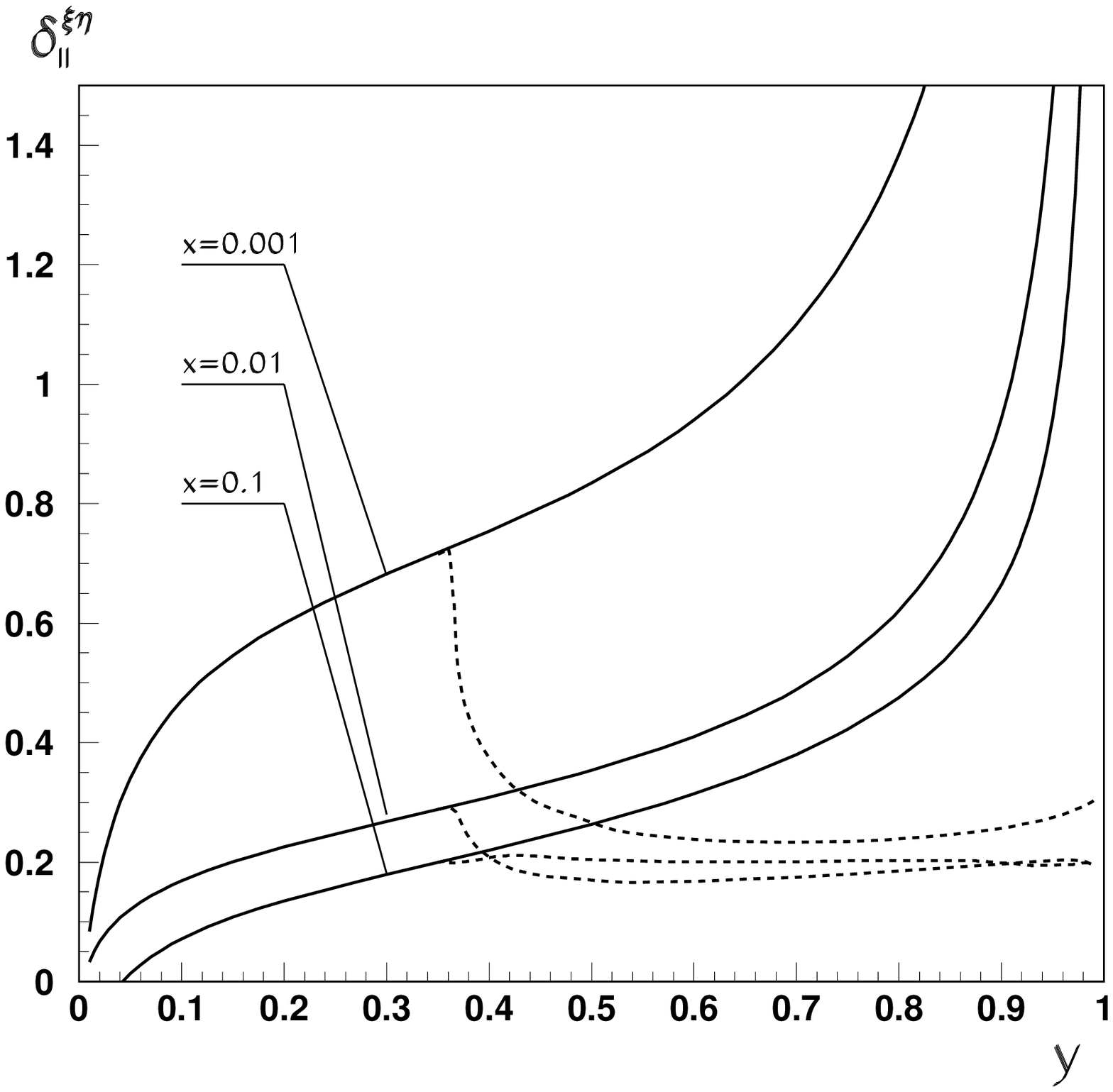}
}
\end{picture}
\\
\begin{picture}(60,60)
\put(-5,0){
\epsfxsize=7cm
\epsfysize=8cm
\epsfbox{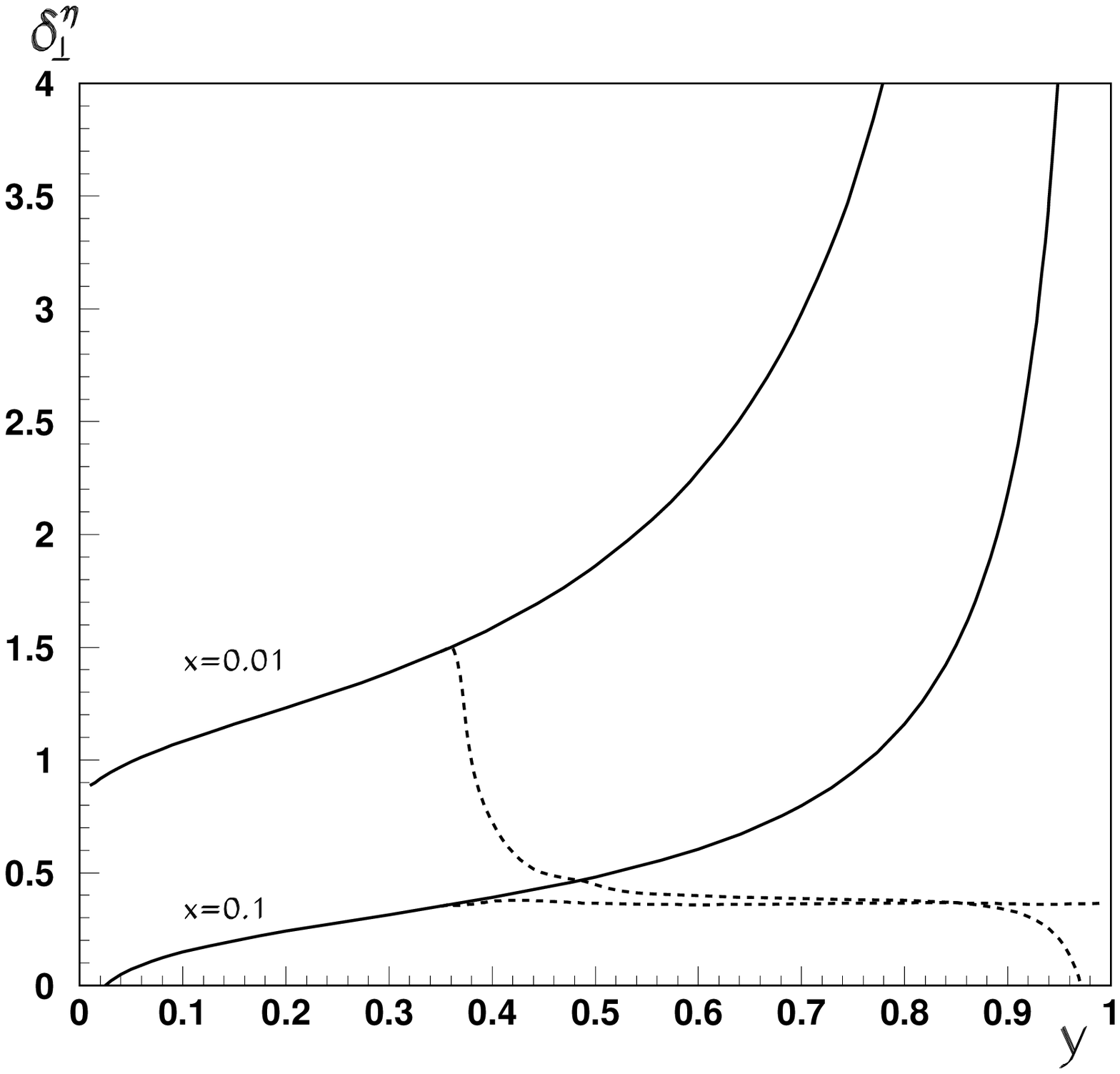}
}
\end{picture}
&
\begin{picture}(60,60)
\put(20,0){
\epsfxsize=7cm
\epsfysize=8cm
\epsfbox{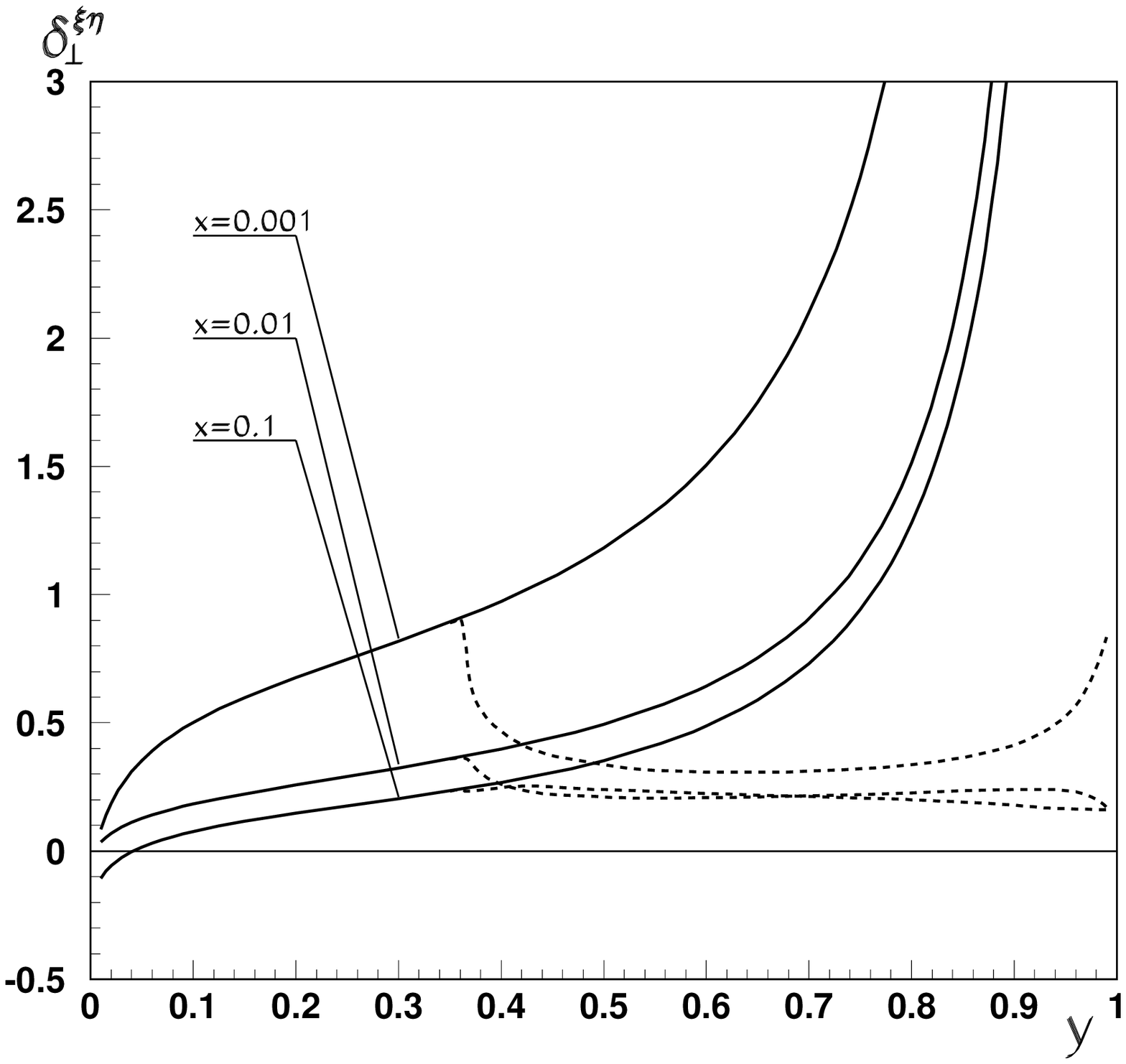}
}
\end{picture}
\end{tabular}
\vspace{-1cm}
\caption{\protect\it
Radiative correction to $\protect\sigma ^{u}$, $\protect\sigma ^{\xi}$,
$\sigma ^{\eta}$ and $\sigma ^{\xi \eta}$ defined in
(\ref{cs}) with (dashed curves) and without cut (full curves). 
 The down indexes $\bot $ and $||$ correspond to longitudinally
and transversely polarized proton beam respectively. 
$\delta ^{\protect\eta}_{||}$ and 
$\delta ^{\protect\eta}_{\protect \bot}$	
are singular for $x=0.001$ due to the Born
cross section $\sigma ^{\eta}$ crossing zero, so  the 
correspondent curves $\delta ^{\eta}$ are rejected. 
}
\label{Fg2}
\end{figure}

In this section the RC to different observable quantities in deep
inelastic electron-proton scattering at collider are studied
numerically. 

The double differential cross section as a function of
the polarization characteristics of the scattering particles can be
presented as the sum of four terms:  
\begin{equation}
\sigma=\sigma^u+P_L\sigma^{\xi}+P_N\sigma^{\eta}
+P_NP_L\sigma^{\xi\eta}, 
\label{cs} 
\end{equation} 
the first of them is an unpolarized cross section and three others
characterize the polarized contributions independent on polarization
degrees.There are no problems with the luminosity measurement in the
current collider experiments, so apart from the usual measurement of
polarized asymmetries the absolute measurement of cross sections with
different polarization configurations of beam and target will be
probably possible in future polarization experiments at collider. 
Besides, now the new methods of data processing, when experimental
information of spin observables is extracted directly from the
polarized part of the cross section \cite{Gagu,Gagu2} are actively
developed. In \cite{Gagu2} it is shown how to separate completely
unpolarized and polarized cross sections from a sample of
experimental data using a special likelihood procedure and a binnig
on polarization degrees. All above mentioned as well as the fact that
RC to asymmetry is always constructed from RC to parts of the cross
section allows to restrict our consideration to numerical studying of
RC to all of the cross sections in the equation (\ref{cs}) and their
combinations.

\begin{figure}
\vspace{2cm}
\hspace{0.5cm}
\unitlength 1mm
\begin{tabular}{cc}
\begin{picture}(60,60)
\put(-5,0){
\epsfxsize=7cm
\epsfysize=8cm
\epsfbox{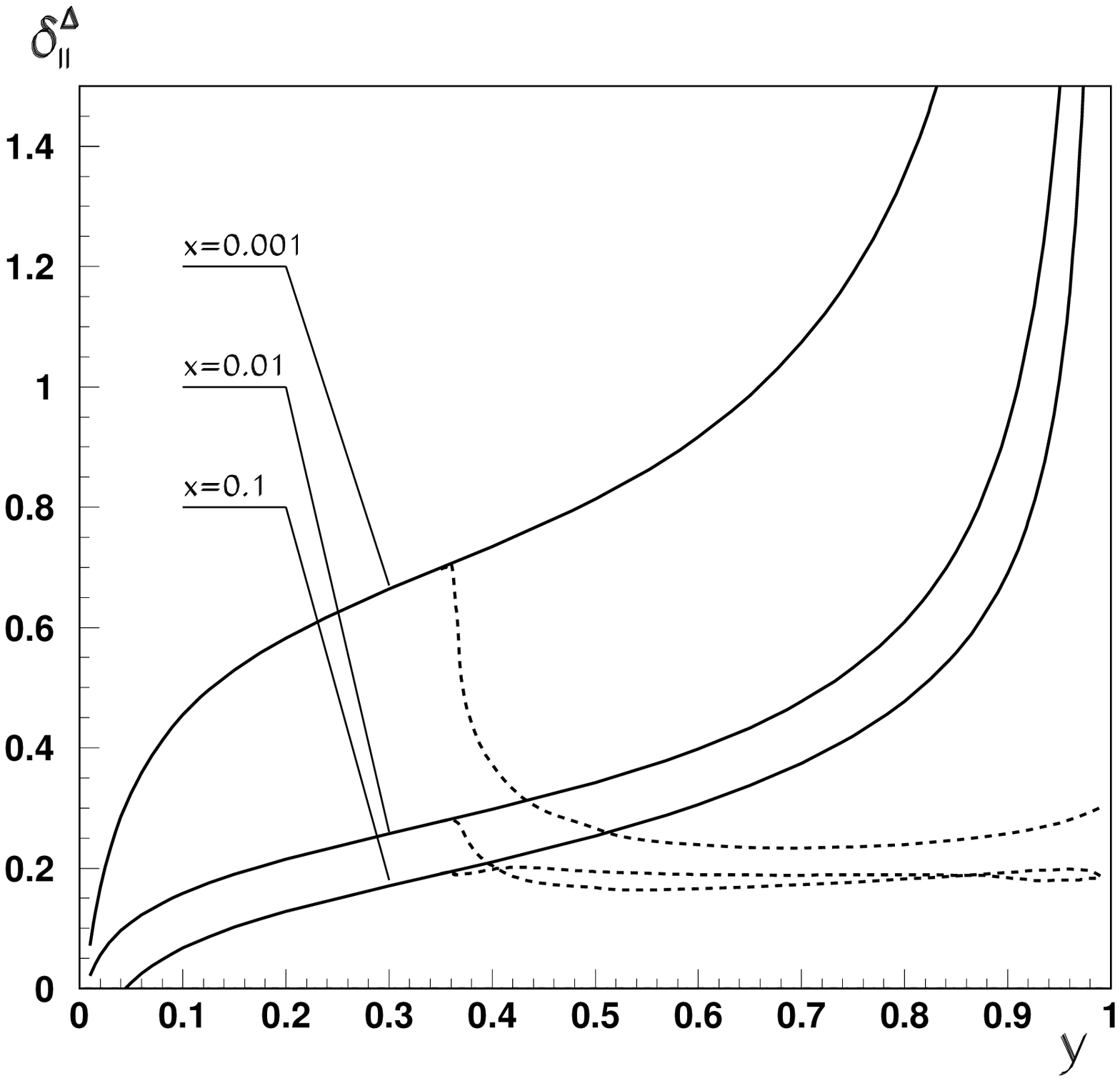}
}
\end{picture}
&
\begin{picture}(60,60)
\put(7,0){
\epsfxsize=7cm
\epsfysize=8cm
\epsfbox{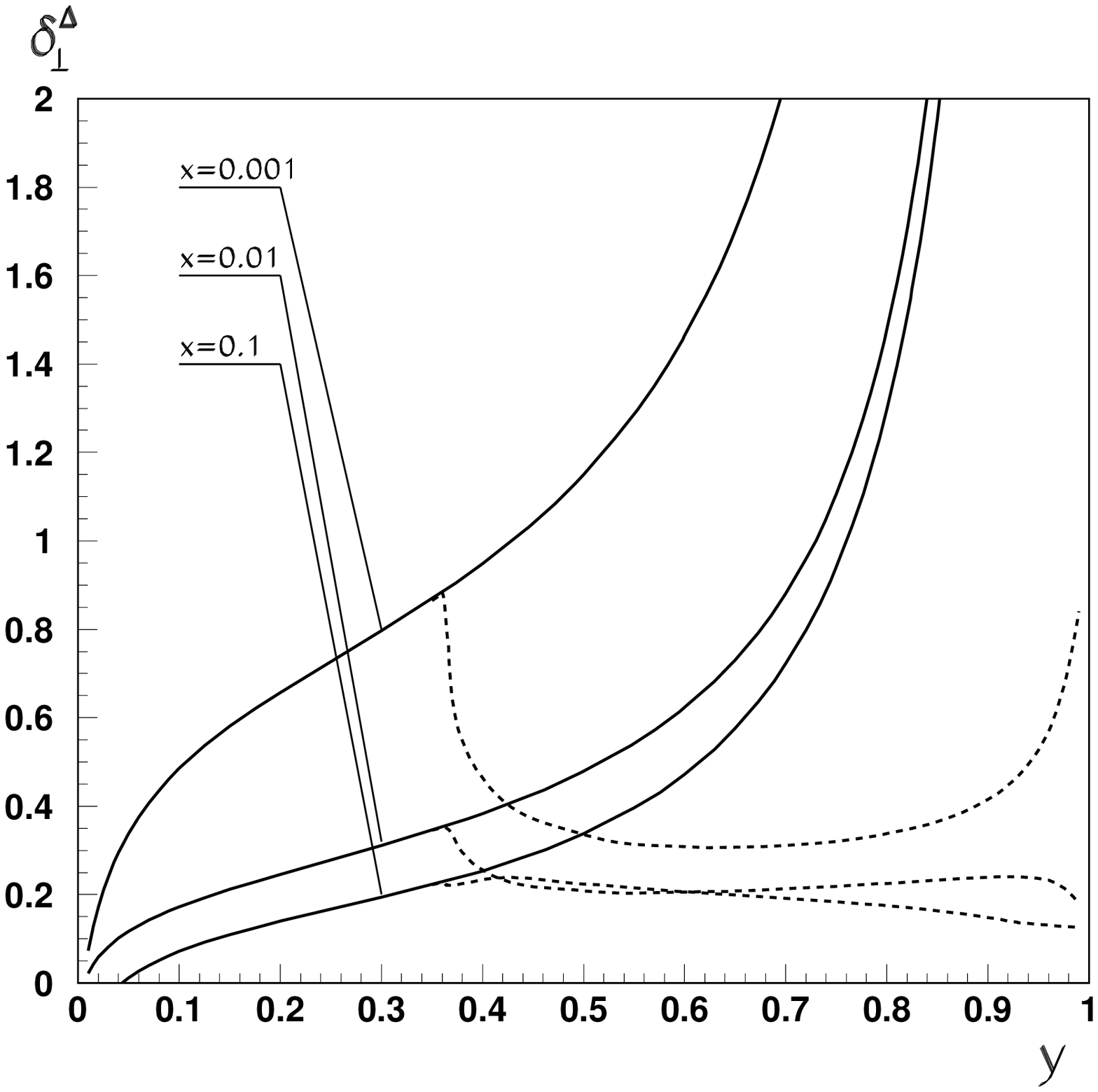}
}
\end{picture}
\end{tabular}
\caption{\protect\it
Radiative correction to
$\Delta \sigma$ defined in 
(\ref{asy}) with (dashed curves) and whithout cut (full curves).
 The down indexes $\protect\bot $ and $||$ correspond to longitudinally
and transversely polarized proton beam respectively.
}
\label{Fg4}
\end{figure}

Radiative correction to these cross sections defined as a ratio
of the cross section including one-loop RC only to the Born one
\begin{equation}
\delta^a=\frac{\sigma_{RC}^a}{\sigma_B^a}
=\frac{\sigma_{tot}^a}{\sigma_B^a}-1,
\qquad
(a=u, \xi , \eta , \xi \eta)
\label{asy0}
\end{equation}
is presented  on figure \ref{Fg2}  as a function of
scaling variables $x$ and $y$. The kinematical region 
corresponds to the  present unpolarization collider experiment at
HERA \cite{hera}. The results were obtained using GRV-parton model
\cite{GRV,GRV2} for the electroweak SF $F_i^{\gamma\;\gamma Z\;Z}$ and
$g_i^{\gamma\;\gamma Z\;Z}$ which are defined by the formulae (23-25) of
ref.\cite{APJ} (see also \cite{DIS}).

 In experiments at collider a detection of a hard photon in calorimeter is
used to reduce radiative effects.  The dashed lines on figure \ref{Fg2}
demonstrate the influence of an experimental cut on the RC. One of the
simplest variant of the cut when events having a radiative photon energy
$E_{\gamma } > 10$GeV are rejected from analysis, is considered only. 

From these plots one can see that the relative RC for polarized part of
the cross section has the same behavior as the unpolarized one:  it goes
up when $y$ tends to kinematical boards ($y\rightarrow 0, y\rightarrow 1$)
and when $x$ goes down. At the same time the correction to polarized parts
can exceed the correction to unpolarized ones several times more.  Usage
of the cut on photon energy does not influence RC in the region of small
$y$ and suppresses RC in the rest of the region. More detailed discussion
of RC at collider with and without experimental cuts can be found in
ref.\cite{AS}. 

\begin{figure}
\vspace{2cm}
\hspace{0.5cm}
\unitlength 1mm
\begin{picture}(60,60)
\put(40,0){
\epsfxsize=7cm
\epsfysize=8cm
\epsfbox{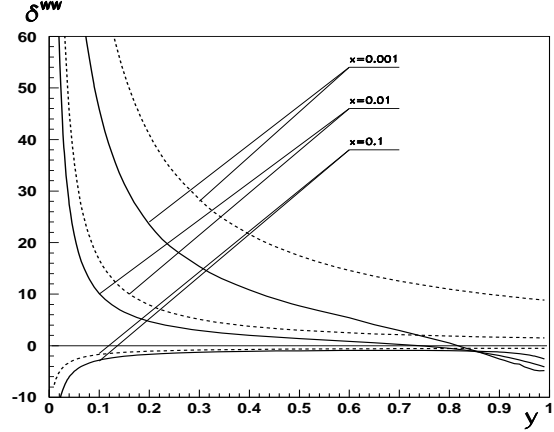}
}
\end{picture}
\vspace{-1cm}
\caption{\protect\it
The quantity $\delta ^{WW} $ on the Born level (dashed curves) and on the  
level of radiative correction (full curves).
}
\label{Fg6}
\end{figure}

The simplest case for absolute measurements in polarization
experiments is the observation of difference of cross sections with
the opposite configuration of the proton spin
\begin{equation}
\Delta \sigma_{||}=\sigma^{\uparrow \Uparrow}
-\sigma^{\uparrow \Downarrow},\qquad
\Delta \sigma_{\bot}=\sigma^{\uparrow \Leftarrow}
-\sigma^{\uparrow \Rightarrow}.
\label{asy}
\end{equation}
The first and the second arrows correspond to the lepton and  proton
polarization degrees equal to $\pm 1$. 
Main contribution to $\Delta \sigma$ comes from the electromagnetic
structure functions $g_1^{\gamma}$ and $g_2^{\gamma}$.
The values of RC to $\Delta \sigma$ defined the same as  (\ref{asy0})
are presented by figure \ref{Fg4}
 for the longitudinally and transversely polarized protons.
This result was obtained using usual
approximation $g_2^{\gamma }=0$. At the same time there is no reason
to be restricted to such a case, especially when we deal with
transversal polarized protons. One of the alternative models is
a well-known Wandzura and Wilcheck's formula
\cite{WW}
\begin{equation}
g_2^{WW}(x,Q^2)=-g_1(x,Q^2)+\int\limits^1_x g_1(\xi,Q^2)\frac {d\xi}{\xi}.
\end{equation}
The quantity characterizing the influence of a model on the 
cross section (\ref{asy}) can be defined as
\begin{equation}
\delta^{WW}=\frac{\Delta \sigma _{\bot}(g_2^{\gamma }=g_2^{WW})-
\Delta \sigma _{\bot}(g_2^{\gamma }=0)}
{\Delta \sigma _{\bot}(g_2^{\gamma }=0)}.
\label{g2}
\end{equation}
Figure \ref{Fg6} shows that the influence is important and
the cross sections (\ref{asy}) calculated with
$g_2^{\gamma }=g_2^{WW}$ and with $g_2^{\gamma }=0$ differ
 in some dozen times in the region of small $x$ and $y$.

Experimental information on electroweak structure functions can be
obtained using the data of electron and positron scattering with
different spin configurations both of leptons and protons.
Correspondent combinations of the cross sections were offered in
ref.\cite{APJ} and discussed in review \cite{DIS}. 
Here we consider two of them, which allow to extract  SF
$g_1^{\gamma Z \; Z}$ and $g_5^{\gamma Z\; Z}$. 
On the Born level they read
\begin{eqnarray}
\fl
\sigma_-^{\uparrow \Uparrow}-
\sigma_-^{\uparrow \Downarrow}+
\sigma_+^{\uparrow \Uparrow}-
\sigma_+^{\uparrow \Downarrow}+	
\sigma_-^{\downarrow \Uparrow}-
\sigma_-^{\downarrow \Downarrow}+
\sigma_+^{\downarrow \Uparrow}-
\sigma_+^{\downarrow \Downarrow}= \nonumber
\\
\lo
=\frac {32\pi\alpha ^2 S}{Q^4} x(2-2y-y^2)
[v^Z\chi g_5^{\gamma Z}+
((a^Z)^2+(v^Z)^2)\chi^2 g_5^{Z}]
\label{asy2}
\end{eqnarray}
and
\begin{eqnarray}
\fl
\sigma_-^{\uparrow \Uparrow}-
\sigma_-^{\uparrow \Downarrow}-
\sigma_+^{\uparrow \Uparrow}-
\sigma_+^{\uparrow \Downarrow}+
\sigma_-^{\downarrow \Uparrow}-
\sigma_-^{\downarrow \Downarrow}-
\sigma_+^{\downarrow \Uparrow}-
\sigma_+^{\downarrow \Downarrow}= \nonumber
\\
\lo
=\frac {32\pi\alpha ^2 S}{Q^4} xy(2-y)
[a^Z\chi g_1^{\gamma Z}+
2v^Za^Z\chi^2 g_1^{Z}],
\label{asy3}
\end{eqnarray}
where the lower index -(+) corresponds to the electron(positron)-proton
scattering.
These expressions were obtained  using model
\cite{DIS, APJ} where 
$g_2^{\gamma Z}=g_4^{\gamma Z}=g_4^{Z}=0$
and $g_3^{\gamma Z, Z}=2xg_5^{\gamma Z, Z}$. 

\begin{figure}
\vspace{2cm}
\hspace{0.5cm}
\unitlength 1mm
\begin{tabular}{cc}
\begin{picture}(60,60)
\put(-5,0){
\epsfxsize=7cm
\epsfysize=8cm
\epsfbox{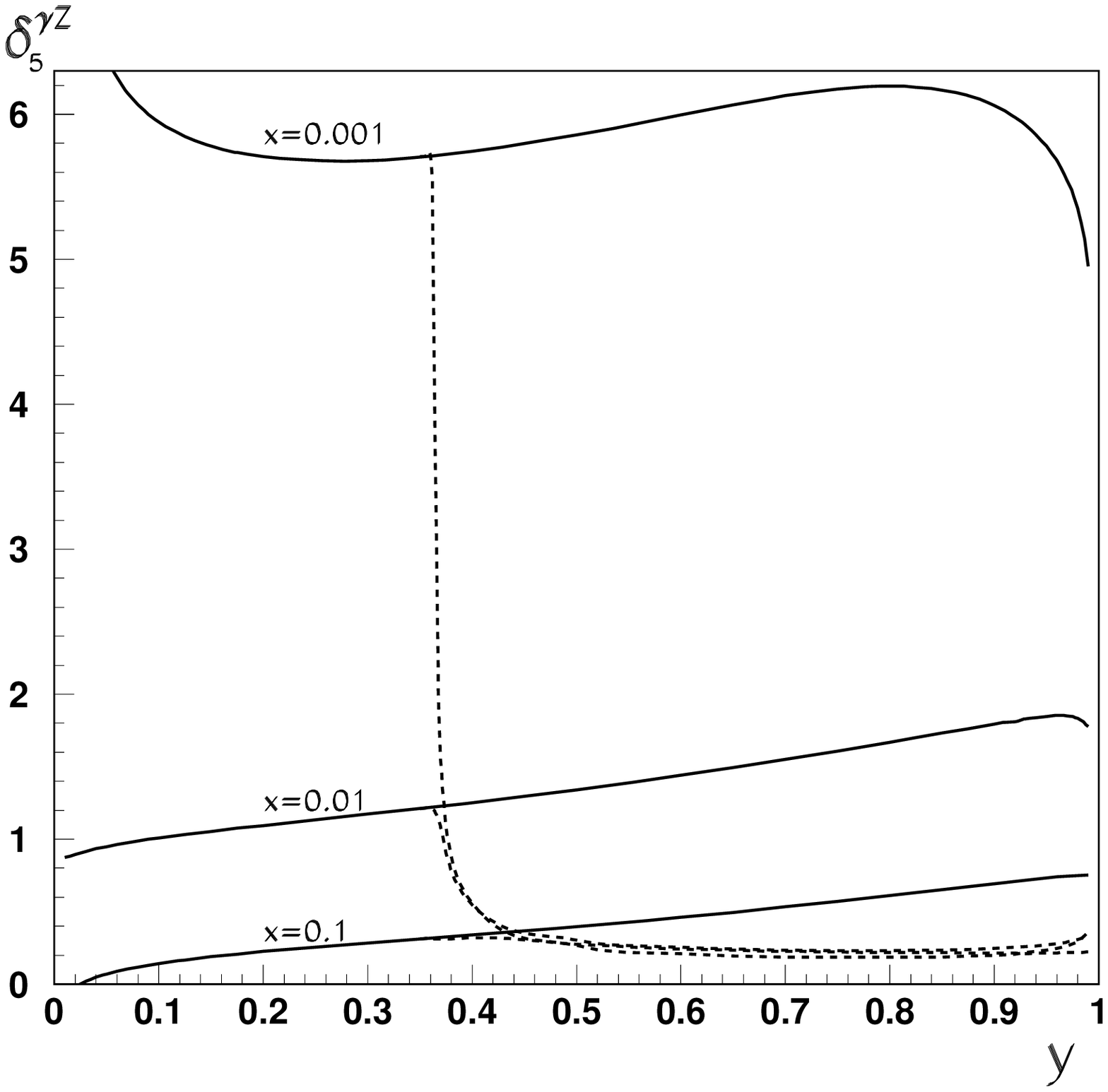}
}
\end{picture}
&
\begin{picture}(60,60)
\put(7,0){
\epsfxsize=7cm
\epsfysize=8cm
\epsfbox{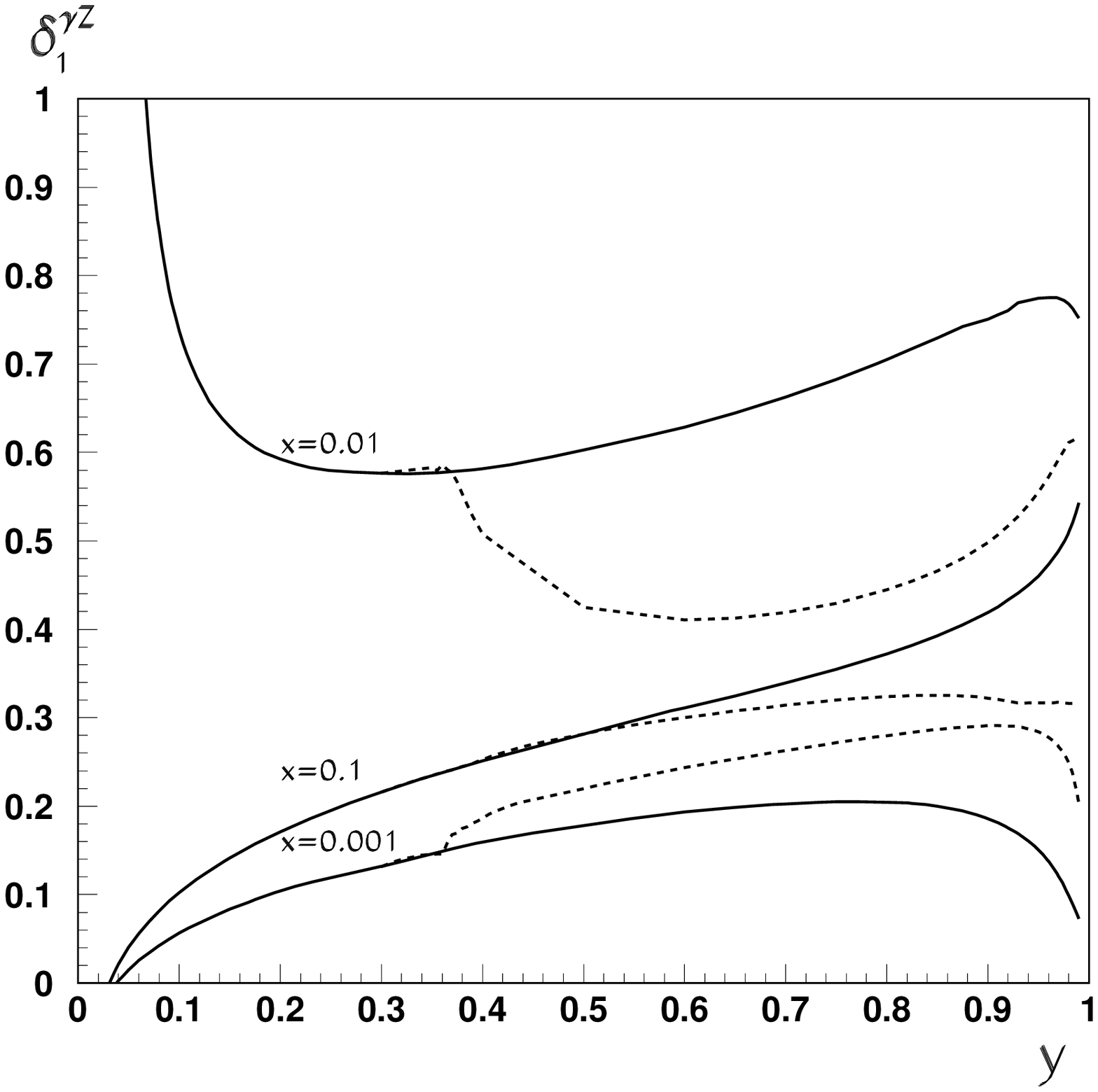}
}
\end{picture}
\end{tabular}
\caption{\protect\it
Radiative correction to quantities  defined in 
(\ref{asy2}) and (\ref{asy3}) with (dashed curves) and
without cut (full curves).
}
\label{Fg7}
\end{figure}

\begin{figure}
\vspace{1cm}
\hspace{.5cm}
\unitlength 1mm
\begin{picture}(60,60)
\put(40,0){
\epsfxsize=7cm
\epsfysize=8cm
\epsfbox{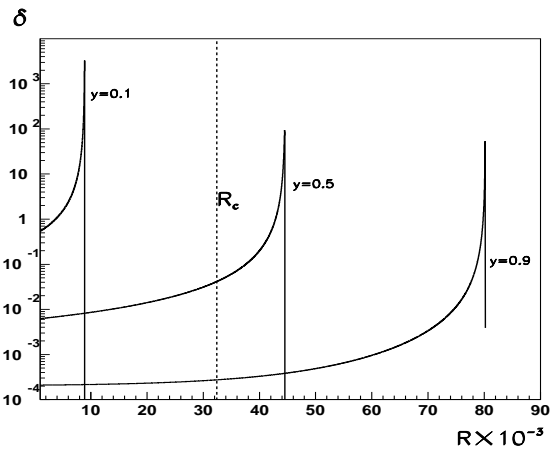}
}
\end{picture}
\vspace{-1cm}
\caption{\protect\it
Integrand of radiative correction
to the cross section (\ref{FR}) 
over $R$ normalized to the Born one in the case of combination
of the cross section (\ref{asy2}).
Dashed line shows the used cut of photon enegy
($E_{\protect\gamma } < 10$ GeV). 
}
\label{Fg5}
\end{figure}

From figure \ref{Fg7} one can see that in the region of small $x$
the radiative correction cross section (\ref{asy2}) exceeds the Born
one in several times. The main effect comes from $\sigma ^F _R$
(\ref{FR}), integrand of which over $R$ for the sum of the cross section
(\ref{asy2}) is sketched on figure \ref{Fg5}. 
It can be understood from the analysis of this figure
that using the experimental cut discussed above
allows essentially to reduce radiative effects for high $y$ and
does not influence the magnitude of RC for $y < 0.3$. 

\section{Conclusion}

Thus we obtained the compact and transparent formulae for the lowest order
model independent electroweak radiative correction within t'Hooft-Feynman
gauge. These formulae can be applied to data processing of experiments at
collider. Numerical analysis of the obtained formulae for kinematics of
collider experiments shows that radiative correction to the unpolarized 
cross section and polarized parts of cross section have the same behavior,
however polarized correction can exceed the unpolarized one in several
times. 
The detection of a hard photon in calorimeter allows to reduce the
radiative effects in the region $y>0.3$.

\section*{Appendix} 

In this appendix the explicit expressions for quantities $\theta_{ij}(\tau
)$ are represented. Due to the factorization of infrared terms all
$\theta_{i1}$ are proportional to Born contributions:
\begin{equation}
\theta_{i1}(\tau )=4F_{IR}\theta^B_i.
\end{equation}

The others $\theta_{ij}(\tau )$ functions are given
\begin{eqnarray}
\fl\theta_{12}=4\tau F_{IR}
\nonumber 
\\ 
\fl\theta_{13}=-2(2F+F_d\tau^2)
\nonumber 
\\
\fl\theta_{22}=(F_{1+}S_xS_p-F_dS_p^2\tau+2m^2F_{2-}S_p
-2(2M^2\tau-S_x)F_{IR})/2M^2
\nonumber 
\\
\fl\theta_{23}=(4FM^2+2F_dM^2\tau^2-F_dS_x\tau-F_{1+}S_p)/2M^2
\nonumber 
\\
\fl\theta_{32}=(F_{1+}S_xQ^2+2m^2F_{2-}Q^2-
m^2F_{2+}S_p\tau+3F_{IR}S_p\tau)/2M^2
\nonumber 
\\
\fl\theta_{33}=(2m^2F_{2-}\tau-F_dS_p\tau^2-2F_{1+}Q^2)/2M^2
\nonumber 
\\
\fl\theta_{34}=-\tau F_{1+}/2M^2
\nonumber 
\\
\fl\theta_{42}=2(\eta {\cal K}\tau(m^2F_{2+}-3F_{IR})-\eta
qF_{1+}Q^2 - 2m^2F_{2-}^{\eta}Q^2)/M
\nonumber 
\\
\fl\theta_{43}=2(\eta {\cal K}F_d\tau^2+2F_{1+}^{\eta}Q^2-
2m^2F_{2-}^{\eta}\tau)/M
\nonumber 
\\
\fl\theta_{44}=2\tau F_{1+}^{\eta}/M
\\
\fl\theta_{62}=(\eta {\cal K}(F_{1+}S_x-2F_dS_p\tau+2m^2F_{2-})
\nonumber 
\\
\lo+\eta
q(F_{1+}S_p+2F_{IR})+2F_{IR}^{\eta}S_x+2m^2F_{2-}^{\eta}S_p)/M
\nonumber 
\\
\fl\theta_{63}=-(\eta {\cal K}F_{1+}+\eta qF_d\tau+
F_d^{\eta}S_x\tau+F_{1+}^{\eta}S_p)/M,
\nonumber 
\end{eqnarray}
and for $i=5,7,8$ :
\begin{eqnarray}
\label{etat}
\fl \theta_{51}=2\eta q\theta_{31}/M,&
\theta_{71}=\eta q\theta_{21}/M,&
\theta_{81}=-\eta q\theta_{11}/M,
\nonumber \\\displaystyle
\fl\theta_{52}=2(\eta q\theta_{32}-\theta^{\eta }_{31})/M,\;\;&
\theta_{72}=(\eta q\theta_{22}-\theta^{\eta }_{21})/M,\;\;&
\theta_{82}=-(\eta q\theta_{12}-\theta^{\eta }_{11})/M,
\nonumber \\\displaystyle
\fl\theta_{53}=2(\eta q\theta_{33}-\theta^{\eta }_{32})/M,&
\theta_{73}=(\eta q\theta_{23}-\theta^{\eta }_{22})/M,&
\theta_{83}=-(\eta q\theta_{13}-\theta^{\eta }_{12})/M,
 \\\displaystyle
\fl\theta_{54}=2(\eta q\theta_{34}-\theta^{\eta }_{33})/M,&
\theta_{74}=-\theta^{\eta }_{23}/M,&
\theta_{84}=\theta^{\eta }_{13}/M,
\nonumber \\\displaystyle
\fl\theta_{55}=-2\theta^{\eta }_{34}/M.
\nonumber
\end{eqnarray}
where ${\cal K}=k_1+k_2$.
The following equalities define the
functions $F$:
\begin{equation}
\eqalign{
F&=\lambda ^{-1/2}_{Q},
\\
F_{IR}&=m^{2}F_{2+}- Q^2F_{d},
\\
F_{d} &=\tau ^{-1} (C^{-1/2}_{2}(\tau )-C^{-1/2}_{1}(\tau ))
\\
F_{1+}&= C^{-1/2}_{2}(\tau ) + C^{-1/2}_{1}(\tau ),
\\
F_{2\pm }&=B_2(\tau ) C^{-3/2}_2(\tau ) \mp B_1(\tau ) C^{-3/2}_1(\tau ),
\\
F_{i}&= - \lambda ^{-3/2}_{Q} B_{1}(\tau ).
}
\end{equation}
Here $\lambda _{Q}=S_x^2+4M^2Q^2$ and
\begin{equation}
\eqalign{
B_{1,2}(\tau )&= - {1\over 2} \pmatrix{\lambda _{Q}\tau \pm
 S_{p}(S_{x}\tau +2Q^2)},
 \\
C_{1}(\tau ) &=(S\tau + Q^2)^{2}+ 4m^{2}(Q^2 + \tau S_x - \tau^{2}M^{2}),
 \\
C_{2}(\tau ) &=(X\tau - Q^2)^{2}+ 4m^{2}(Q^2 + \tau S_x - \tau^{2}M^{2}).
}
\label{bc}
\end {equation}
For all $i$,$j$ $\theta^{\eta}_{ij}$ in (\ref{etat}) are defined as:
\begin{equation}
\theta^{\eta}_{ij}=
\theta _{ij}(F_{All}\rightarrow F^{\eta}_{All}),
\label{th12} 
\end{equation}
where
\begin{equation}
\eqalign{
2F^{\eta}&=F(r_{\eta}-\tau s_{\eta})+2F_{i}s_{\eta},
\\
2F_{2+}^{\eta}&=
(2F_{1+}+\tau F_{2-})s_{\eta}+F_{2+}r_{\eta},
\\
2F_{2-}^{\eta}&=(2
F_{d}+F_{2+})\tau s_{\eta}+F_{2-}r_{\eta},
\\
2F_{d}^{\eta}&=F_{1+}s_{\eta}+F_{d}r_{\eta},
\\
2F_{1+}^{\eta}&=(4F+\tau ^2F_d)s_{\eta}+F_{1+}r_{\eta}.
}
\end{equation}
The quantities
\begin{equation}
s_{\eta}=a_{\eta}+b_{\eta},\;
r_{\eta}=\tau (a_{\eta}-b_{\eta})
+2c_{\eta}
\end{equation}
are the combination of coefficients of the polarization vector
$\eta$ expansion over the basis (see also Appendix A of ref. \cite{ASh})
\begin{equation}
\eta=2(
a_{\eta} k_1 +
b_{\eta} k_2 +
c_{\eta} p).
\label{th11}
\end{equation}
For example in the case of the longitudinally polarized beam they read: 
\begin{equation}
a_{\eta}=
\frac {M}{\sqrt{\lambda _s}},
\;\; 
b_{\eta}=0, 
\;\; 
c_{\eta}=
-\frac {S}{2M\sqrt{\lambda _s}}. 
\label{th113}
\end{equation}


\vspace{13mm}
\Bibliography{99}
\bibitem {EMC}
Ashman J \etal 1988 {\it Phys. Lett.} {\bf B206} 364
\bibitem {KSh}
Kukhto T and Shumeiko N 1983 {\it Nucl. Phys.} {\bf B219} 412
\bibitem {ASh}
  Akushevich I and Shumeiko N, 1994 {\it J. Phys.} {\bf G20} 513
\bibitem {SSh}
Soroko A and Shumeiko N 1989 {\it Yad. Phys.} {\bf 49} 1348
\bibitem {Tim}
Shumeiko N and Timoshin S 1991
 {\it J. Phys.} {\bf G17} 1145
\bibitem {HERA}
Feltesse J, Sch\" afer A 1995/96 {\it Future Physics at HERA}
(DESY, Hamburg: Proceedings of the Workshop) p~760
\bibitem{AISh}
Akushevich I, Ilyichev A and Shumeiko N 1995 {\it  Phys. At. Nucl.}
{\bf 58} 1919
\bibitem {Bardin}
Bardin D, Blumlein J, Christova P and Kalinovskaya L 1997
{\it  Nucl. Phys. } {\bf B506} 295
\bibitem {DIS}
Anselmino M, Efremov A and Leader  E 1995 {\it Phys. Rep.} {\bf 261} 1
\bibitem {Bl}
Blumlein J and Kochelev N 1997 {\it Nucl. Phys.} {\bf B498} 285 
\bibitem {P20}
Akushevich I, Ilyichev A, Shumeiko N, Soroko A and Tolkachev A 1997 {\it
Comp. Phys. Commun.} {\bf 104} 201
\bibitem {Holl}
Hollik W 1990 {\it Fortschr. Phys.} {\bf 38} 165
\bibitem {BHS}
B\"ohm M, Spiesberger H and Hollik W 1986 {\it Fortschr. Phys.} {\bf 34}
687
\bibitem {hera}
Adloff C, \etal 1997 {\it Z. Phys.} {\bf C74 } 196
\bibitem{Gagu}
Gagunashvili N 1994  {\it Nucl. Instr. Meth.}, {\bf A343}  606 
\bibitem{Gagu2}
Gagunashvili N \etal 1996 {\it JINR--E1--96--483} (Dubna) 
{\it to be published in Nucl. Instr.  Meth.~A}
\bibitem {GRV}
Gl\"uck M, Reya E and Vogt A 1992 {\it Z. Phys. } {\bf C53 } 127 
\bibitem {GRV2}
Gl\"uck M, Reya E, Stratmann M and Vogelsang W 1996 {\it Phys. Rev. } {\bf
D53 } 4775 
\bibitem {APJ}
Anselmino M, Gambino P, Kalinovski J 1994 {\it Z. Phys. } {\bf C64 } 267
\bibitem {AS}
Akushevich I, Spiesberger H 1995/96 {\it Future Physics at HERA}
(DESY, Hamburg: Proceedings of the Workshop)
p~1007
\bibitem {WW}
Wandzura W and  Wilczek F 1977 {\it Phys. Lett.} {\bf B172} 195
\endbib
\end{document}